\documentclass{elsart}
\usepackage{natbib,graphicx,amssymb}
\journal{New Astronomy Reviews, accepted}

\newcommand{\oversim}[2]{\protect{\mbox{\lower0.5ex\vbox{%
  \baselineskip=0pt\lineskip=0.2ex
  \ialign{$\mathsurround=0pt #1\hfil##\hfil$\crcr#2\crcr\sim\crcr}}}}}
\newcommand{\simgreat}{\mbox{$\,\mathrel{\mathpalette\oversim>}\,$}} % >~ sign
\newcommand{\simless} {\mbox{$\,\mathrel{\mathpalette\oversim<}\,$}} % <~ sign
\begin{document}

\begin{frontmatter}

\title{Massive Stars: Their Birth Sites and Distribution}

\author[1,2]{Pavel Kroupa}
\ead{pavel@astrophysik.uni-kiel.de}

\address[1]{
Insitut f\"ur Theoretische Physik und Astrophysik, Universit\"at
Kiel, D-24098 Kiel, Germany}

\address[2]{Heisenberg Fellow}

%--------------------------------------------------------------------------
\begin{abstract}
The stellar IMF has been found to be an invariant Salpeter power-law
($\alpha=2.35$) above about $1\,M_\odot$, but at the same time a
massive star typically has more than one companion. This constrains
the possible formation scenarios of massive stars, but also implies
that the true, binary-star corrected stellar IMF could be
significantly steeper than Salpeter, $\alpha> 2.7$.  A significant
fraction of all OB stars are found relatively far from potential birth
sites which is most probably a result of dynamical ejections from
cores of binary-rich star clusters. Such cores form rapidly due to
dynamical mass segregation, or they are primordial. Probably all OB
stars thus form in stellar clusters together with low-mass stars, and
they have a rather devastating effect on the embedded cluster by
rapidly driving out the remaining gas leaving expanding OB
associations and bound star clusters.  The distributed population of
OB stars has a measured IMF with $\alpha\approx 4$, which however,
does not necessarily constitute a different physical mode for isolated
star formation. A steep field-star IMF is obtained naturally because
stars form in clusters which are distributed according to a power-law
cluster mass function.
\end{abstract}


\end{frontmatter}

%------------------------------------------------------------------------
\section{Introduction}
\label{sec:intro}

Stars with masses $m> 10\,M_\odot$ are very much the power engines of
galaxies leading brief but brilliant lives. They are the major energy
sources for galactic atmospheres and rapidly inject heavy elements
into them where they die thus changing galactic weather patterns and
thereby playing a major role in regulating star formation.
Understanding their distribution, both by mass and spatially, and
their birth is therefore of underlying importance for understanding
the physical state and chemical content of the interstellar medium and
the morphological appearance of galaxies.

This contribution gives an overview of the initial mass function
(IMF), of the pairing properties and of the spatial distribution of
massive stars, as well as their impact on the birth cloud and survival
of star clusters. This subject is also reviewed by Massey (1998,
distribution and IMF), Schaerer (2003, IMF), Zinnecker (2003,
multiple-star properties) and Garay \& Lizano (1999, environment and
formation).

%------------------------------------------------------------------------
\section{The IMF}
\label{sec:imf}

The initial mass function, $\xi(m)$, is the distribution by mass of
unevolved, zero-age stars, $dN=\xi(m)\,dm=\xi_{\rm L}(m)\,d{\rm
log}_{10}m$ being the number of stars in the mass interval $m,m+dm$
and log$_{10}m$, log$_{10}m+d$log$_{10}m$, respectively, where
$\xi_{\rm L}(m)=(m\,{\rm ln}10)\,\xi(m)$ is the ``logarithmic IMF''.
Inferring masses of stars is in general difficult, because they cannot
be weighed. Rather, their luminosity is converted to a mass. This can
be tricky though, because optical luminosities do not provide a
one-to-one map to the stellar mass, since for example a 4~Myr old,
$40\,M_\odot$-star is brighter with $M_{\rm V}=-6.6$ than a zero-age,
$120\,M_\odot$-star which has $M_{\rm V}=-6.2$. Optical photometry is
not well suited for measuring the masses of massive stars because they
emit most of their radiation in the UV, and spectral classification
together with photometry is required (Massey 1998, 2003).  If the age,
effective temperature, composition and rotational angular momentum
vector of a single star are known, then the mass can be calculated
given theoretical stellar models. Typically only the luminosity is
known rather well, and the effective temperature or spectral type is
determined from the star's colour or from spectral analysis. The other
quantities are typically not well known and mass estimates are
therefore limited to an uncertainty given by the changes in mass
induced for changes in compositions and rotation rates.
Stellar-evolution models also need to be improved. For example, Massey
(2003) points out that observed red super-giants (RSGs) lie at far too
low effective temperatures than are reached by modern evolution
models, and that the observed relative numbers of RSGs and WR stars
are not reproduced theoretically.

Placing the stars of a population in the HR diagram, their masses can
be inferred from a comparison with theoretical stellar evolution
tracks. Massey and collaborators have been studying the IMF in OB
associations with different metallicities and densities $\simless
1\,M_\odot$/pc$^3$, and found for $m\simgreat$a~few$\,M_\odot$ that
$\xi(m)\propto m^{-\alpha}$ with $\alpha=2.3\pm0.1$ ($N_{\rm OB}=1$
association in the Small Magellanic Cloud with metallicity $Z=0.002$),
$\alpha=2.3\pm0.1$ ($N_{\rm OB}=10$ in the Large Magellanic Cloud,
LMC, with $Z=0.008$), and $\alpha=2.1\pm0.1$ ($N_{\rm OB}=13$ with
$Z=0.02$ in the Milky Way, MW).  The stellar IMF is thus, to a very
good approximation, a Salpeter (1955) power-law, and there is no
significant dependence on $Z$ for $Z\simgreat 0.002$. This rather
amazing result is perplexing, because the metallicity defines the
cooling rate of a molecular cloud core and thus its fragmentation
behaviour, and a dependency of the IMF on $Z$ should therefore
exist. Perhaps the dependency will become apparent when young
populations with $Z<0.002$ are probed. Also, there does not appear to
be a significant dependency on the density of a population. The dense
($\approx 10^4\,M_\odot$/pc$^3$) and 1--3~Myr old massive star-burst
cluster R136 in the 30~Doradus star-forming region in the LMC contains
about 39~O3 stars which is more than known to be contained in the rest
of the MW, LMC and SMC combined, and again $\alpha=2.35\pm0.14$ is
found for $m\simgreat 2.8\,M_\odot$ (Massey \& Hunter 1998; Massey
1998).

The mass of the most massive star in a population, $m_{\rm max}$, also
appears to be independent of $Z\simgreat 0.002$ (Massey 1998). This is
an important finding because it suggests that radiation pressure on
dust grains may not be a major limiting process during the assembly of
massive stars.  If the stellar IMF is a Salpeter power-law with a
flattening at low masses (see below), then R136, which has a stellar
mass of roughly $10^{5}\,M_\odot$, should contain $m>750\,M_\odot$
stars (provided they can form) and any star cluster more massive than
$8000\,M_\odot$ should contain stars as massive as $200\,M_\odot$
(Elmegreen 2000; Weidner \& Kroupa 2003). Because stars more massive
than $150\,M_\odot$ have not been found in any population, Weidner \&
Kroupa infer that a fundamental maximum stellar mass, $m_{\rm max*}$,
must exist, such that $m_{\rm max}\le m_{\rm max*}\approx150\,M_\odot$
independent of the richness of the population. Stars more massive than
about $150\,M_\odot$ cannot form.  This conclusion, based on
statistics, does not hold true if the stellar IMF were steeper with
$\alpha>2.8$ for $m>1\,M_\odot$, because then stars more massive than
about $150\,M_\odot$ would be too rare to occur even in R136 and even
if they can form.

Indeed, the true stellar IMF (the IMF obtained by counting all
individual stars in a population) is probably significantly steeper
than the measured Salpeter value (\S~\ref{sec:mult}).  For
completeness it should be mentioned that the true stellar IMF is a
Salpeter power-law, $\alpha=2.3\pm0.3$ in the mass range
$0.5\,M_\odot$--a~few$\,M_\odot$. It flattens to $\alpha_1=1.3\pm0.5$
for $0.08\le m/M_\odot < 0.5$ and even further to $\alpha_0=0.3\pm0.7$
below $0.08\,M_\odot$ (Reid et al. 1997; Chabrier 2003; Kroupa 2001,
2002). This IMF is referred to in the following as the ``standard
IMF''.

%------------------------------------------------------------------------
\section{Multiples}
\label{sec:mult}

Defining the companion-star fraction as $CSF = (N_B + 2\,N_T + 3\,N_Q
+ \ldots)/(N_S + N_B + N_T + N_Q + \ldots)$, where the nominator is
the number of companions in $N_{\rm sys}=N_S + N_B + N_T + N_Q +
\ldots$ systems or primaries, and $N_S, N_B, N_T, N_Q$ is the number of
single, binary, triple and quadruple systems, respectively, then
massive stars typically have $CSF\approx 1.5$ (Preibisch, Weigelt \&
Zinnecker 2001; Zinnecker 2003), while T~Tauri stars have $CSF\approx
1$ (Duch{\^ e}ne 1999) and late-type main-sequence stars typically have
$CSF\approx 0.5$.

A particularly well studied population of massive stars is the Orion
Nebula Cluster (ONC). The ONC is about 1~Myr old and contains in total
between 5000~and $10^4$~stars (Hillenbrand 1997; Kroupa 2000), and is
still very compact and most probably expanding as a result of very
recent gas blow-out (Kroupa, Aarseth \& Hurley 2001). It contains
today 27~OB ``stars'' (Hillenbrand 1997), of which the central
Trapezium is composed of the four systems $\theta^1$~Ori A,B,C,D.
Bispectrum speckle interferometry of 13~OB stars by Preibisch et
al. (1999) in the ONC together with additional data shows that the
``Trapezium'' consists of at least 11~stars with a combined mass of
about $100\,M_\odot$.  The low-mass cluster members have $CSF\approx
0.5$ while the OB~stars have a significantly larger $CSF\approx 1.5$.
The companion separations of the surveyed systems range from 0.13~to
460~AU.  Preibisch et al. conclude that on average O~stars have~1.5
to~4 companions, and that the distribution of companions is peaked
towards low-mass stars.  Duch{\^ e}ne et al. (2001) survey the
occurrence of companions within distances of 200--3000~AU around 60~OB
stars in the massive young cluster NGC~6611. For mass-ratios larger
than~0.1, $CSF>1$, and the companion mass-distribution is found to be
consistent with random sampling from the IMF.  Such systematic and
detailed observational surveys are needed for other systems containing
massive stars, and Zinnecker (2003) overviews the available results.

The binary-star properties of massive stars are a key constraint on
their formation mechanism (next section). But one immediate
implication is that the high $CSF$ implies that the true stellar IMF
cannot be a Salpeter power-law, because for each massive primary there
are typically a few less-massive companions. Detailed corrections to
the true $\alpha$ are not available yet, although work is in progress
(Weidner \& Kroupa, in preparation). However, a guess can be obtained
by noting that for a 100\% binary fraction ($CSF=1$) and masses picked
randomly from the IMF an observed power-law IMF with $\alpha_{\rm
obs}=2.35$ (\S~\ref{sec:imf}) has a true underlying $\alpha\approx
2.7$ (Sagar \& Richtler 1991). Since $CFS>1$ for OB stars,
$\alpha>2.7$.

%------------------------------------------------------------------------
\section{Clusters and OB associations}
\label{sec:cl}

About 50\% of all OB stars are found within associations, and most of
the remaining ones are in clusters.  It can thus be deduced that most
if not all OB stars form in embedded clusters.  Assuming a standard
IMF with $\alpha=2.35$ or $\alpha=2.7$ ($m>1\,M_\odot$) implies that a
single star with a mass of $20\,M_\odot$ is associated with a cluster
of 810 or 3400 stars, respectively. Indeed, observations of rich young
clusters show that massive stars and low-mass stars form together; a
separate star-formation mode favouring massive stars is not evident
(Kroupa 2002). This will be re-addressed in \S~\ref{sec:isol}.
Low-mass clusters that contain fewer than a few tens of late-type
O~stars loose their gas within a cluster-dynamical time once the
massive stars ``ignite'' causing rapid expansion of the stellar
system. The binding energy of the embedded cluster, $E_{\rm
bin}=G\,M_{\rm st + g}^2/R = 9\times 10^{48}$~erg
($9\times10^{50}$~erg) for a star$+$gas mass of $10^4\,M_\odot$
($10^5\,M_\odot$) and cluster radius $R=1$~pc with, respectively,
crossing times (further below) of 0.3~Myr (0.1~Myr). According to the
stellar-evolution models of Maeder (1990), a~$15\,M_\odot$
($85\,M_\odot$) main-sequence star injects $3\times10^{50}$~erg
($3\times10^{51}$~erg) of radiation and wind energy per 0.1~Myr into
the cloud. Self-consistent radiation-transport hydrodynamics
computations such as presented by Freyer, Hensler \& Yorke (2003) are
needed to address this issue in detail, but the above estimates
suggest that stellar feedback alone can inject enough energy on a
time-scale shorter than the dynamical time to unbind the cloud.  The
resulting evolutionary sequence UC\,HII region $\longrightarrow$ HII
region and the relevant physical processes are described by Garay \&
Linzano (1999).  A molecular cloud region producing a number of such
clusters will thus emerge as an expanding OB association. Massive
embedded clusters, such as R136, cannot expel their gas within a
dynamical time owing to the depth of the potential well (Goodwin 1997)
and consequently survive longer in a denser configuration allowing
more time for energy re-distribution among the cluster stars thereby
forming bound star clusters that are surrounded by expanding stellar
populations (Kroupa \& Boily 2002; Boily \& Kroupa 2003a,b).

Within the clusters, the massive stars are typically centrally
concentrated. Examples are the massive R136 cluster, which has been
found to be mass-segregated (Selman et al. 1999; Bosch et al. 1999)
and the low-mass ONC (Hillenbrand 1997), which hosts the Trapezium
system at its centre (\S~\ref{sec:mult}).  It is not clear yet if the
observed degree of mass segregation in clusters younger than a few~Myr
is dynamical or primordial in origin.  An idealised star cluster
consisting of single stars with different masses that is initially in
dynamical equilibrium evolves by seeking a new equilibrium state that
can never be achieved. An encounter between two unequal stars in the
cluster potential imparts kinetic energy to the less-massive star
which therefore becomes less bound to the cluster at the expense of
the more massive one which consequently sinks towards the potential
minimum thereby getting more bound to the cluster. As a result it
picks-up kinetic energy again and can again exchange its energy with
another less-massive star. The process continues until other
astrophysical effects such as stellar collisions or mass loss from the
cluster core through stellar winds begin to dominate, leaving an
expanded low-mass stellar population with an increasingly bound core
of massive stars.  The time-scale for energy equipartition in an
idealised cluster has been estimated by Spitzer (1977, p.74) to be
$t_{\rm msegr} \approx (m_{\rm av}/m_{\rm m})\,t_{\rm relax}$. Here
$m_{\rm av}$ is the average stellar mass ($m_{\rm av}\approx
0.4\,M_\odot$) and $m_{\rm m}$ is the mass of the massive star taken
to be $m_{\rm m}=20\,M_\odot$ for the examples below.  The relaxation
time, which is the time for significant energy redistribution to occur
in a cluster (e.g. Binney \& Tremaine (1987), $t_{\rm relax} \approx
(N/8\,{\rm ln}N)\,t_{\rm cross}$, and $t_{\rm cross} = 2\,R / \sigma$
is the crossing time of a typical star through the cluster which has a
characteristic radius $R$ and a velocity dispersion $\sigma =
(G\,M/R)^{0.5}$, where $G=0.0045\,{\rm pc}^3/(M_\odot\,{\rm Myr}^2)$
is the gravitational constant. For example, for the post-gas-expulsion
cluster R136 with $M_{\rm st}\approx10^5\,M_\odot$ and $R\approx1$~pc
$t_{\rm cross}\approx0.1$~Myr, $t_{\rm relax}\approx250$~Myr and
$t_{\rm msegr}\approx 5$~Myr, while for the pre-gas-expulsion ONC with
$M_{\rm st+g}=10^4\,M_\odot$ and $R\approx 0.4$~pc, $t_{\rm
cross}\approx0.08$~Myr, $t_{\rm relax}\approx11$~Myr ($10^4$~stars)
and $t_{\rm msegr}\approx 0.2$~Myr. Thus, in both cases the
equipartition time-scale is comparable to the age of the clusters
precluding firm conclusions on the nature of the observed mass
segregation. The estimate $t_{\rm msegr}$ is very crude because once
the massive stars dominate the central region there will not be enough
low-mass stars to carry away the energy and the process must
slow. Self-consistent stellar-dynamical calculations are needed to
address this problem in detail.  Bonnell \& Davies (1998) argue that
the ONC is too young for mass segregation to have been able to proceed
sufficiently far to account for the observed signature. Their
modelling relies on using a force-softened stellar-dynamical code to
solve the equations of motion, and assumes all stars to be single.
Applying a direct summation code with force regularisation, thus
avoiding the need for force softening, and assuming a high binary
fraction and more massive and denser initial (pre-gas-expulsion)
cluster models, Kroupa et al. (2001) found mass segregation to proceed
sufficiently far to perhaps account for the observed mass segregation
in the densest model of the ONC (Kroupa 2002). This model has an
initial central density of $10^{5.8}$~stars/pc$^3$.

Knowing if the observed mass segregation is primordial is important
for constraining formation-theories of massive stars.  Radiation
pressure on infalling gas with an inter-stellar dust content is
sufficient to halt mass accretion for stars more massive than about
$10\,M_\odot$ unless the accretion rate surpasses
$10^{-3}\,M_\odot$/yr and the dust abundance is suppressed (Wolfire \&
Cassinelli 1987), and Bonnell, Bate \& Zinnecker (1998) suggest that
collisions between interme\-diate-mass proto-stars in accreting
cluster cores may lead to runaway collisional behaviour as the central
potential deepens and more low-angular momentum gas is accreted
causing further core contraction and so on. This model naturally leads
to a high $CSF$ for massive stars, since not all the proto-stars will
merge, especially so since during the final stage gas infall is
reversed due to the energetic feedback from the central massive
star(s), immediately leading to core expansion (Vine \& Bonnell 2003).
The accretion-induced proto-stellar collision scenario is supported by
Preibisch et al. (1999) on the basis of the multiple-star properties
of the massive stars in the ONC and would also lead to a natural
explanation of the observed mass segregation in very young clusters,
but it requires very high central densities, $10^8$~stars/pc$^3$. This
may be compatible with some ultra-compact HII regions that have
dimensions of about 0.1~pc (Zinnecker 2003), but the processes
occurring within these remain unreachable observationally.
Observational evidence from star-forming cloud regions and thus
extremely young systems with ages of a few~0.1~Myr do not appear to
support this picture entirely. A pre-cluster cloud region with
dimensions of not more than a few~pc often shows complex morphology
with massive stars spread throughout the volume (e.g. Tieftrunk et
al. 1998 for the W3 star-forming region containing more than 10~HII
regions within a region spanning 2~pc; Alves \& Homeier 2003 for the
W49 region which contains more than 100~O stars, about 30 being within
a 6~pc diameter region).  In some cases massive disks have been
detected around massive proto-stars perpendicular to their outflows
(e.g. NGC7\,538S is an embedded $\approx 40\,M_\odot$ star with a
30\,000~AU, $400\,M_\odot$ rotating disk, Sandell, Wright \& Forster
2003). Accretion from a disk lessens the radiation-pressure problem
and allows very massive stars to form because accretion occurs first
onto the disk and then onto the protostar in its equatorial plane,
allowing thermal energy to escape polewards (Jijina \& Adams 1996;
Yorke \& Sonnhalter 2002). Massive molecular outflows are also
observed emanating from massive proto-stars, implying that they may
form like low-mass stars albeit from much denser and warmer cloud
cores (Garay \& Lizano 1999; McKee \& Tan 2003). That massive stars
may form through proto-stellar accretion together with cloud
fragmentation is supported by Duch{\^e}ne et al. (2001) on the basis
of the wide-binary properties of massive stars in M~16, and would be
consistent with the very early scattered distribution of massive
proto-stars within forming clusters, but would require dynamical mass
segregation to proceed swiftly enough to account for the observed mass
segregation in young, post-gas-expulsion clusters.

However they form, cores of massive stars are unstable because a
small$-N$ system decays in about $N-$crossing times. For example, for
the ONC the central Trapezium system has a radius $R_{\rm C}\approx
0.02$~pc, mass $M_{\rm C}\approx 100\,M_\odot$ so that $t_{\rm
cross,C}\approx 0.008$~Myr and the core with 10~stars should decay in
about 0.08~Myr. Since the ONC has an age of about 1~Myr, the central
Trapezium should have decayed long ago, especially if the
pre-gas-expulsion density was higher. Even if we allow 0.5~Myr for the
Trapezium to form through dynamical mass segregation, it should have
decayed a few times by now.  This suggests that the Trapezium should
be in the final stages of decay, and that the ONC core may have hosted
more massive stars than are observed within the Trapezium today. The
other massive stars would have been ejected. The two most massive
stars in the ONC, $\theta^1$~Ori~C and $\theta^2$~Ori~A, are found to
have large proper motions (van Altena et al. 1988) perhaps due to
on-going break-up of the cluster core.

%------------------------------------------------------------------------
\section{Isolated O stars}
\label{sec:isol}

Of all O~stars 10--25\% are runaways with speed $v\simgreat40$~km/s,
while only 2\% of all B~stars and only 0.1--0.2\% of all A~stars are
runaways. Fast O~stars have a binary proportion 2--4 times lower than
low-velocity OB stars (Gies \& Bolton 1986). Clearly these data have
rather important implications for the chemo-dynamical evolution of
galaxies, because they indicate that a sizeable fraction of massive
stars explode relatively far from their birth-sites.  Given typical
birth velocities of massive stars of a few~km/s it has been argued
that they cannot drift, within their short lifetimes, to the positions
where they are found.  In the LMC a number of O3 stars with life-times
as short as 2~Myr are found 100~pc or more from viable birth
sites. The existens of such outlying massive stars poses a challenge,
and it has been argued that such cases may constitute examples of an
exotic birth-mode which produces only one massive star and little
else, which may be possible if isolated molecular clouds have a stiff
equation of state (Spaans \& Silk 2000; Li, Klessen \& McLow
2003). The IMF of the distributed population of OB stars has been
measured, $\alpha\approx4\pm0.5$ (Massey 1998), and may indicate a
variation of the IMF with star-forming conditions (but see the end of
this section). However, the invariance of the stellar IMF among OB
associations with different metallicity (\S~\ref{sec:imf}) does not
appear to support this possibility.

As concluded in \S~\ref{sec:cl} massive stars prefer to reside in
dynamically unstable cluster cores, and violent ejections from these
may well explain most if not all of the outliers. The sense of the
effect would, qualitatively, give the correct trend with stellar mass,
because the most-massive stars would be more prone to violent
ejections by virtue of their core-residence than the less massive
B~and the even less massive A~stars which are not observed to form
cluster cores with such high binding energies. The trend of runaways
with stellar mass cannot, on the other hand, be obtained if a primary
star explodes as a supernova thereby freeing its companion which
leaves with the circular (because of tidal circularisation of the
short-period orbit prior to the explosion) orbital velocity (a
few~100~km/s, e.g. Tauris \& Takens 1998). The problem with this
scenario, which undoubtedly leads to some runaways (Portegies Zwart
2000), is that the mass-ratio distribution of massive binaries
(\S~\ref{sec:mult}) implies that less-massive companions (B~and
A~stars) ought to be expelled with large velocities more often than
massive ones because they are more frequent (\S~\ref{sec:mult}).

In an elegant analysis, Clarke \& Pringle (1992) deduce certain
characteristics that the massive stars need to have in order for the
statistical properties of the observed runaways to be explained as
being dynamically ejected stars from cluster cores. They conclude that
the massive stars need to form in small groups ($N\simless 20$)
containing massive binaries with mass-ratios biased towards
unity. These groups need to be severely underrepresented in low-mass
stars. Modern binary-star data (\S~\ref{sec:mult}) do not appear to be
consistent with these constraints, although for example a deficit of
low-mass stars has been verified for the core of the ONC (Hillenbrand
1997).  Sophisticated stellar-dynamical calculations need to be
performed in order to understand the core processes in detail, and to
verify the conjecture of Clarke \& Pringle. Such work has been in
progress, and for example Zinnecker (2003) reports numerical
experiments of clusters with ($CSF\ge0$) and realistic IMFs, whereby
the binaries are constructed by randomly pairing stars from the IMF in
the mass range $0.01-50\,M_\odot$ and there is no initial mass
segregation. This null hypothesis leads to far too many massive
primaries having secondaries with a mass ratio $<0.1$ than are
observed, but the massive stars rapidly segregate to the cluster
centres, exchange secondaries for more massive ones, and for the
binary-rich cases up to 40\% of all stars with $m\ge8\,M_\odot$ are
found outside twice the tidal radius of their cluster by the time they
explode. This thus appears to support the dynamical-ejection scenario
for a large fraction of OB runaways, but much more detailed work is
necessary for firmer confirmation.

Because binary--binary encounters dominate the production of runaway
stars in clusters containing binaries, the fundamental interaction
reaction is of the type
binary$+$binary$\longrightarrow$triple(unstable or
stable)$+$single$_{\rm ej}$ ($B+B\longrightarrow T^{\rm
u\,or\,s}+S_{\rm ej}$) and if the triple is unstable then $T^{\rm u}
\longrightarrow B+S_{\rm ej}$ follows within typically a few dynamical
times of the unstable triple, although the decay occurs stochastically
and cannot be predicted given the chaotic nature of the general
3--body Kepler problem. Leonard (1991) has performed a vast number of
$B+B$ scattering experiments for different binary-star mass ratios in
order to quantify the reaction cross sections and thus the probability
of outcome. The asymptotic ejection velocities that can be reached
after an infinite number of $B+B$ events are $v_{\rm ej} \le v_{\rm
esc}$ for an ejected low-mass star, where $v_{\rm esc}$ is the escape
velocity from the surface of the most massive star involved in the
quadruple reaction, and $v_{\rm ej} \le 0.5\,v_{\rm esc}$ for an
ejected star with a mass equal to the most massive star in the
reaction. For example, $v_{\rm esc}=1400$~km/s for a $60\,M_\odot$
star so that $v_{\rm ej}=700$~km/s for another $60\,M_\odot$ star.

Quantitative demonstration that a given OB star has been ejected is
difficult and time-consuming, because its orbit in the Galactic
potential needs to be calculated backwards in time in order to
identify the possible young cluster of origin. This has been achieved
beautifully in a few cases (Hoogerwerf et al. 2001; Allen \& Kinman
2003), but some OB~stars appear to be located so far from any
star-forming sites that they could not have propagated to their
present positions given realistic velocities and birth sites near to
the Galactic disk. Nevertheless, careful scrutiny (e.g. Ramspeck,
Heber \& Moehler 2001) typically verifies that ejection is viable in
most cases.  Extreme ejection velocities may be reached in the
reaction $B+bh\longrightarrow (bh+S)+S_{\rm ej}$, where $S_{\rm ej}$
can attain a velocity as large as 4000~km/s if $bh$ is a
$10^6\,M_\odot$ black hole (Hills 1988).

Returning to the field-star IMF, Kroupa \& Weidner (2003) show that it
must be steeper ($\alpha_{\rm field} > \alpha$ for $m>1\,M_\odot$)
than the true, binary-star-corrected IMF which has $\alpha\simgreat
2.7$ (\S~\ref{sec:mult}). Stars form in embedded clusters most of
which disperse within a few~100~Myr and which are distributed
according to a cluster mass function which is found to be well
represented by a power-law with index $\beta\approx 2.2$ (Lada \& Lada
2003; Hunter et al. 2003). Low-mass clusters contain no massive stars,
so that the field-star IMF contains fewer massive stars per low-mass
star than the population within an individual well-populated cluster
does. The result is that $\alpha_{\rm field}\approx3.5$ for
$\alpha=2.7$, coming close to the field-star IMF determined by Massey
and collaborators (beginning of this section).

%------------------------------------------------------------------------
\section{Conclusions}
\label{sec:concs}

The birth of each massive stars is associated with violent
gas-dynamical processes and the emergence of a cluster containing
thousands of low-mass stars.  The measured initial mass function (IMF)
of massive stars in clusters is found to be a universal Salpeter
power-law ($\alpha=2.35$).  The maximum stellar mass appears to be
near $150\,M_\odot$ and does not depend on metallicity for $Z\simgreat
0.002$. The limit near $150\,M_\odot$ may constitute a physical mass
limit, or it may constitute a statistical limit beyond which stars are
simply too rare to be found even in the most massive clusters. The
latter can only be the case if the true, binary-star corrected IMF has
$\alpha>2.7$, which in fact is probably the case because most massive
stars are in higher-order multiple systems with a mass-ratio
distribution that is consistent with random pairing from the IMF and
not with a preference towards similar-mass companions. This may
perhaps be a result of the coalescence of intermediate proto-stars in
accreting cluster cores, but formation through accretion from a
massive, unstable disk may also be a possible formation scenario.  The
field-star IMF, which is the relevant distribution function for energy
feedback and metal deposition on galactic scales, is again steeper
than this ($\alpha_{\rm field}\simgreat 3.5$) because stars form in a
distribution of star clusters with different masses.  Despite their
short lifes, a significant fraction of all massive stars are situated
far from potential birth sites. In contrast, intermediate stars do not
spread as far, and the reason is most probably that massive stars
reside in dynamically unstable binary-rich cores in clusters from
which they are expelled rapidly. At this stage there is no conclusive
evidence for a special mode of star formation which favours production
of massive stars in isolated regions of galactic atmospheres.

Key issues that await further scrutiny are improvement of stellar
models, measurement of the IMF in very-low metallicity regions,
quantification of the multiple-star properties in dependence of
metallicity, the theory of stellar-dynamical ejections and the
formation of cluster cores through dynamical mass segregation, the
detailed physical processes involved in expelling residual gas from an
embedded cluster, and its reaction to the rapid mass loss in
dependence of cluster mass. 

%-----------------------------------------------------
I thank the organising committee for the invitation to present this
overview at beautiful Kloster Seeon in Bavaria, and I acknowledge
support through grants KR1635/3 and KR1635/4.

\begin{small}
\begin{description}

\itemsep=-0.5mm

\item Allen, C., Kinman T. D. (2003), in Sixth Pacific Rim Conference
  on Stellar Astrophysics - A tribute to Helmut, Kluwer Publ., in
  press (astro-ph/0212272)

\item Alves, J. and Homeier, N. (2003), {\it Ap.J.L.}, {\bf 589}, L45 

\item Binney, J. and Tremaine S. (1987),
  Galactic Dynamics, Princeton University Press

\item Boily, C.~M. and Kroupa, P. (2003a), {\it M.N.R.A.S.}, {\bf 338}, 665 

\item Boily, C.~M. and Kroupa, P. (2003b), {\it M.N.R.A.S.}, {\bf 338}, 673

\item Bonnell, I.~A. and Davies, M.~B. (1998), {\it M.N.R.A.S.}, 
  {\bf 295}, 691 

\item Bonnell, I.~A., Bate, M.~R., and Zinnecker, H. (1998), {\it
  M.N.R.A.S.}, {\bf 298}, 93

\item Bosch, G., Terlevich, R., Melnick, J., \& Selman, F. (1999), {\it
  A\&AS}, {\bf 137}, 21 

\item Chabrier, G. (2003), {\it PASP}, {\bf 115}, 763

\item Clarke, C.~J. and Pringle, J.~E. (1992), {\it M.N.R.A.S.}, {\bf
  255}, 423

\item Duch{\^ e}ne, G. (1999), {\it A\&A}, {\bf 341}, 547 

\item Duch{\^ e}ne, G., Simon, T., Eisl{\"o}ffel, 
  J., Bouvier, J. (2001), {\it A\&A}, {\bf 379}, 147 

\item Elmegreen, B.~G. (2000), {\it Ap.J.}, {\bf 539}, 342

\item Freyer, T., Hensler, G., Yorke, H. W. (2003), {\it Ap.J.}, 594, 888

\item Garay, G. and Lizano, S. (1999), {\it PASP}, {\bf 111}, 1049 

\item Gies, D.~R. and Bolton, C.~T. (1986), {\it Ap.J.S}, {\bf 61}, 419 

\item Goodwin, S.~P. (1997), {\it M.N.R.A.S.}, {\bf 284}, 785 

\item Hillenbrand, L.~A. (1997), {\it AJ}, {\bf 113}, 1733 

\item Hills, J.~G. (1988), {\it Nature}, {\bf 331}, 687 

\item Hoogerwerf, R., de Bruijne, J.~H.~J. and de Zeeuw, P.~T. (2001),
  {\it A\&A}, {\bf 365}, 49

\item Hunter D. A., Elmegreen B. G., Dupuy T. J. and Mortonson M. (2003)
  {\it AJ}, 126, 1836

\item Jijina, J. and Adams, F.~C. (1996), {\it Ap.J.}, {\bf 462}, 874 

\item Kroupa, P. (2000), {\it New Astronomy}, {\bf 4}, 615 

\item Kroupa, P. (2001), {\it M.N.R.A.S.}, {\bf 322}, 231

\item Kroupa, P. (2002), {\it Science}, {\bf 295}, 82 

\item Kroupa, P. and Boily, C.~M. (2002), {\it M.N.R.A.S.}, {\bf 336}, 1188 

\item Kroupa, P. and Weidner C. (2003), {\it Ap.J.}, in press
  (astro-ph/0308356)

\item Kroupa, P., Aarseth, S., \& Hurley, J. (2001), {\it M.N.R.A.S.},
   {\bf 321}, 699

\item Lada, C. J. and Lada E. A. (2003), {\it ARAA}, {\bf 41}, in press
  (astro-ph/0301540)

\item Leonard, P.~J.~T. (1991), {\it AJ}, {\bf 101}, 562

\item Li, Y., Klessen, R.~S. and Mac Low, M. (2003), {\it Ap.J.}, {\bf
  592}, 975

\item Maeder, A. (1990), {\it A\&AS}, {\bf 84}, 139 

\item Massey, P. (1998), {\em The Stellar Initial Mass Function
    (38th Herstmonceux Conference)}, G.~Gilmore, D.~Howell (eds), ASP
    Conference Series, 142, 17

\item Massey, P. (2003), in May 2003 STScI Symp. "The Local Group as
  Astrophysical Laboratory", in press (astro-ph/0307531)

\item Massey, P. and Hunter, D.~A. (1998), {\it Ap.J.}, {\bf 493}, 180 

\item McKee, C.~F. and Tan, J.~C. (2003), {\it Ap.J.}, {\bf 585}, 850 

%\item Megeath, S.~T., Herter, T., Beichman, C., Gautier, N., Hester,
%  J.~J., Rayner, J., and Shupe, D.  (1996), {\it A\&A}, {\bf 307}, 775

\item Reid, I.~N., Gizis, J.~E., \& Hawley, S.~L. (2002), {\it AJ},
   {\bf 124}, 2721

\item Portegies Zwart, S.~F. (2000), {\it Ap.J.}, {\bf 544}, 437 

\item Preibisch, T., 
   Balega, Y., Hofmann, K., Weigelt, G., \& Zinnecker, H. (1999), {\it
   New Astronomy}, {\bf 4}, 531 

\item Preibisch, T., Weigelt, G., \& Zinnecker, H. (2001), in IAU
   Symposium, {\bf 200}, 69

\item Ramspeck, M., Heber, U., \& Moehler, S. (2001), {\it A\&A}, {\bf
  378}, 907

\item Sagar, R.~\& Richtler, T. (1991), {\it A\&A}, {\bf 250}, 324

\item Salpeter, E.~E. (1955), {\it Ap.J.}, {\bf 121}, 161 

\item Sandell, G., Wright, M. and Forster, J.~R. (2003), {\it Ap.J.L},
  {\bf 590}, L45

%\item Scalo, J.~M. (1986), {\it Fundamentals of Cosmic Physics}, {\bf 11}, 1 

\item Schaerer, D. (2003), in IAU Symp. 212, K. A. van der Hucht,
  A. Herrero, C. Esteban, eds, p.642 (astro-ph/0208227)

\item Selman, F., Melnick, J., Bosch, G., and Terlevich, R. (1999),
{\it A\&A}, {\bf 347}, 532

\item Spaans, M., Silk, J., (2000), {\it Ap.J.}, {\bf 538}, 115

\item Spitzer L. (1987), Dynamical Evolution of Globular
  Clusters, Princeton University Press

\item Tauris, T.~M. and Takens, R.~J. (1998), {\it A\&A}, {\bf 330}, 1047 

\item Tieftrunk, A.~R., Megeath, S.~T., Wilson, T.~L., \& Rayner,
  J.~T. (1998), {\it A\&A}, {\bf 336}, 991

\item van Altena, W.~F., Lee, J.~T., Lee, J.-F., Lu, P.~K., and
  Upgren, A.~R. (1988), {\it AJ}, {\bf 95}, 1744

\item Vine, S.~G. and Bonnell, I.~A. (2003), {\it M.N.R.A.S.}, {\bf 342}, 314 

\item Weidner, C., \& Kroupa, P. (2003), {\it M.N.R.A.S.}, accepted
	(astro-ph/0310860)

\item Wolfire, M.~G. and Cassinelli, J.~P. (1987), {\it Ap.J.}, {\bf
  319}, 850

\item Yorke, H.~W. and Sonnhalter, C. (2002), {\it Ap.J.}, {\bf 569}, 846 

\item Zinnecker, H. (2003), in IAU Symp. 212, K. A. van der Hucht,
  A. Herrero, C. Esteban, eds, in press (astro-ph/0301078)

%
%\bibitem[Chieffi, Limongi and Straniero (1998)]{cls} Chieffi, A.,
%Limongi, M. and Straniero O. (1998), {\it Ap.J.}, {\bf 502},~737

\end{description}
\end{small}
%\end{thebibliography}

\end{document}